\documentclass[journal]{IEEEtran}
\IEEEoverridecommandlockouts
\usepackage{cite}
\usepackage{amsmath}
\usepackage{amsmath,amssymb,amsfonts}
\usepackage{algorithmic}
\usepackage{graphicx}
\usepackage{textcomp}
\usepackage{amsfonts}
\usepackage{graphicx}
\graphicspath{{figures/}}
\usepackage{tikz, gincltex}
\usepackage{caption}
\usepackage{pgfplots}\pgfplotsset{compat=1.9}
\usepackage{amsmath}
\usepackage{array}
\usepackage{graphics}
\usepackage{array, booktabs, caption}
\usepackage{yfonts}
\usepackage{subfig}
\usepackage{mathtools}

\usepackage{ulem}

\def\BibTeX{{\rm B\kern-.05em{\sc i\kern-.025em b}\kern-.08em
    T\kern-.1667em\lower.7ex\hbox{E}\kern-.125emX}}
\begin{document}

\title{Trust-aware Safe Control for Autonomous Navigation: Estimation of System-to-human Trust for Trust-adaptive Control Barrier Functions}

\author{
Saad Ejaz and Masaki Inoue,~\IEEEmembership{Member,~IEEE}
\thanks{S. Ejaz and M. Inoue are with the Department
of Applied Physics and Physico-Informatics, Keio University,
Yokohama 223-8522, Japan, e-mail: saadejazz@keio.jp, minoue@appi.keio.ac.jp.}
\thanks{This work was supported by Grant-in-Aid for Scientific Research (B), No. 20H02173 JSPS.}
}

\DeclarePairedDelimiter\abs{\lvert}{\rvert}
\DeclarePairedDelimiter\norm{\lVert}{\rVert}

\maketitle

\begin{abstract}
A trust-aware safe control system for autonomous navigation in the presence of humans, specifically pedestrians, is presented. The system combines model predictive control (MPC) with control barrier functions (CBFs) and trust estimation to ensure safe and reliable navigation in complex environments. Pedestrian trust values are computed based on features, extracted from camera sensor images, such as mutual eye contact and smartphone usage. These trust values are integrated into the MPC controller's CBF constraints, allowing the autonomous vehicle to make informed decisions considering pedestrian behavior. Simulations conducted in the CARLA driving simulator demonstrate the feasibility and effectiveness of the proposed system, showcasing more conservative behaviour around inattentive pedestrians and vice versa. The results highlight the practicality of the system in real-world applications, providing a promising approach to enhance the safety and reliability of autonomous navigation systems, especially self-driving vehicles.
\end{abstract}

\section{Introduction}
The rapid development of autonomous technology has increased the deployment of robots in human-populated environments. While this brings benefits in efficiency, productivity, and convenience it raises concerns about safety. Establishing trust between humans and autonomous systems is crucial for integrating these systems into society \cite{beer2014toward}. Extensive studies have explored the importance of trust in autonomous systems \cite{adams2003trust}, measured individual trust in robots \cite{schaefer2013perception}, and proposed future directions for analyzing trust \cite{nahavandi2019trust}.

While increased transparency and personalization to improve \textit{human-to-system trust} in autonomous system is extensively researched \cite{sun2020exploring, akash2020human, akash2019improving}, works exploring \textit{system-to-human trust} are less common. Designing safe systems for human-populated environments is challenging due to unpredictable human behavior. Humans are non-deterministic, making it difficult to design systems that can anticipate human behavior. Therefore, safety-critical systems often prioritize caution over performance, leading to overly conservative control designs. 

This papers explores trust-aware safe navigation that models system's trust on humans in the environment to adopt assertive policies near trusted humans while considering extra safety for humans with lower trust. Autonomous driving scenarios are the focus of this research, with pedestrian trust estimated based on behavioral indicators extracted from perceived images of pedestrians. Trust estimates are integrated into control barrier function (CBF) \cite{ames2019control, ames2017cbfqp, amesadaptive} constraints  in model predictive control (MPC) framework to guarantee safety, while making navigation decisions considering the relative trust values. The proposed system can be applied to enhance safety and efficiency in various applications in addition to autonomous driving, including service robots, multi-agent systems, and human-robot collaborative systems.

Trust-aware control design has gained significant attention in recent years to improve the safety and efficiency of navigation systems. The work of \cite{sabbir2023trust} utilizes a resilient control and coordination scheme by using trust-based search and robust scheduling for smooth traffic in a network of connected and automated vehicles. Trust in this case was computed as confidence in the conformance or violation of a set of constraints, tracked by a central coordinator. Another work \cite{ozkan2022trustaware} proposes a non-linear MPC for longitudinal motion planning in a car following interaction with bounds on trust, computed from the difference in the actual plan of the autonomous car and the plan perceived by the human driver following it. Such trust modelling was also employed by \cite{zahedi2023trust} for human-robot interaction scenarios.

CBFs are commonly used in numerous navigation systems with safety governing parameters dependent on trust. The works by \cite{cosner2023learning, lyu2022responsibilityassociated, lyu2023risk} use responsibility allocations as a CBF parameter for autonomous driving applications. Responsibility allocation, a concept similar to trust, dictates the contribution of each autonomous agent in a multi-agent system towards ensuring safety. This is computed as relative social value orientation (R-SVO) between pairwise agents in \cite{lyu2022responsibilityassociated}, and from risk estimation of agents using risk maps inspired by CBFs in \cite{lyu2023risk}. The closest work to this paper \cite{parwana2022trust} utilizes trust metric based on distance and approach angle to identify the nature of other agents in a multi-agent system. This is employed in a rate-tunable CBF framework to yield less conservative control policies.

The utilization of CBFs in trust-aware navigation applications in literature reinforces their efficacy. This work distinguishes itself from prior trust-based CBF research by combining MPC with trust-based CBF constraints, resulting in enhanced safety assurances through the consideration over a future time horizon. Moreover, with regards to trust estimation, most works in this domain either estimate trust for non-human agents (invoking predictable behavior or predetermined natures) \cite{valtazanos2011intent, parwana2022trust, sabbir2023trust, brito2021learning}, or consider a form of human-to-system trust \cite{fooladi2021bayesian, hu2021trust, zahedi2023trust, ozkan2022trustaware}. In contrast, this work focuses on the quantification of unpredictable human (specifically pedestrians) behavior to determine the autonomous system's trust on humans.

Furthermore, estimating pedestrian trust is closely linked with determining inattention or distraction. Existing works on quantifying pedestrian inattention place a heavy emphasis on detecting smartphone usage by pedestrians. Pedestrians immersed in their devices while navigating through or near busy traffic leads to reduced awareness to roadside events, which has been linked to accidents \cite{frej2022smartphone, lin2017impact}. This makes smartphone engagement an important indicator of pedestrian inattention and hence trust. Prior works in this domain include \cite{shinmura2015pedestrian, shinmura2017recognition} which use HOG features-based SVM classifier to detect smartphone engagement while walking; \cite{rangesh2016pedestrians, rangesh2018vehicles} which incorporate pose information, smartphone location, and gaze estimation to identify distraction; and \cite{saenz2021detecting, hatay2021learning} which utilize video pairs to detect smartphone engagement via a deep learning framework. However, the datasets employed for smartphone engagement detection are not public and lack variety in camera angles, lighting conditions, and backgrounds. Furthermore, works exploring pedestrian distraction indicators other than smartphone usage employ conepts such as analysis of working memory during walking \cite{uemura2016estimating}, danger estimation \cite{kusakari2020deep}, and detection of mutual eye contact \cite{belkada2021pedestrians, hata2022detection}.

The majority of existing works in pedestrian distraction/inattention estimation focuses on a single behavior to indicate the pedestrian distraction. Moreover, most image-based classifiers treat each image as an independent entity, lacking the incorporation of time-series data. Inspired from prior works introduced before, this paper devises a comprehensive formulation of trust by incorporating multiple indicators of trust, proposing models to quantify some of those indicators, and exploring the concept of trust dynamics.
The contributions of this paper are as follows:

\begin{enumerate}
\item A formulation of system-to-pedestrian trust from multiple pedestrian behavior correlated to inattention. 
\item Developing a dataset and model for smartphone usage detection as an indicator of pedestrian distraction.
\item An MPC controller with discrete time trust-adaptive CBF constraints
\item Simulation of the proposed trust estimation and control system on an autonomous driving simulator - CARLA.
\end{enumerate}
The remainder of the paper is structured as follows. Details of our proposed methodology are divided into two sections: trust estimation and trust-aware control are addressed in Section \ref{section-trust} and Section \ref{section-control}, respectively. Each of these section introduce the respective formulations with numerical simulations. Section \ref{section-simulation} features simulation of simple autonomous driving scenarios in CARLA using both the trust estimation and trust-aware navigation methodologies introduced in the paper. 

\section{Trust Estimation}
\label{section-trust}
Images are a valuable data source for estimating trust indicators, specifically behavior traits, which are quantified by trait scores denoted by $s$ in this work. Pedestrian images offer insights into trust indicators like smartphone usage, gaze direction, age, and gait analysis that can serve as reliable measures of pedestrians' distraction levels. We consider $N_s$ trust indicating traits (with respective trait scores) for $N_p$ pedestrians. Then, we propose a scoring system to compute the total score $\mathcal{S}_j(t)$ for pedestrian $j$ at timestep $t$ as the linear combination of $N_s$ trait scores.
\begin{align}
    \mathcal{S}_j(t) = \sum_{i=1}^{N_s}\rho_i s_{i, j}(t) \quad \forall \; j \in J,
\label{lin_comb}
\end{align}
where $J$ denotes the set of pedestrian indices given by $\{1,2,\hdots,N_p\}$, and $s_{i, j} \in [\,0, 1\,]$ denotes the score associated with the $i$th behavior trait and $\rho_i$ denotes the constant coefficients of the linear combination satisfying $\sum_{i=1}^{N_s}\rho_i = 1$.
The trait score indicates the trustworthiness with regards to each specific behavior trait; for example, a higher trait score for gait analysis signifies a more stable and reliable gait, contributing more to the overall trust value, and vice versa.
The resulting total score $\mathcal{S}_j$ ranges from 0 to 1, indicating the relative level of attentiveness, with  1 indicating that the autonomous agent fully trusts the pedestrian $j$.

Furthermore, the total score for each tracked pedestrian is aggregated over time by considering a series of captured images, to compute their trust value $\tau$. The aggregation dynamics of trust $\tau_j$ for perceived pedestrian $j$ is modeled by 
\begin{equation}
\tau_j(t+1) = \mbox{sat}_{[0,1]}(\alpha\tau_j(t) + \beta\mathcal{S}_j(t + 1)),
\label{agg_trust}
\end{equation}
where $\mbox{sat}_{M}(v)$ is the saturation function, which maps value $v$ onto set $M$; $\alpha$ and $\beta$ are constants such that $\alpha,\, \beta \in [\,0, 1\,]$; and the initial trust $\tau_j(0)$ is given by
\begin{equation}
\tau_j(0) = \beta_0\mathcal{S}_j(0), \nonumber
\end{equation}
for some positive constant $\beta_0 \in (\,0, 1\,]$.

To track pedestrians between images, the pre-trained tracking-enabled ShuffleNet model \cite{zhang2018shufflenet} from the OpenPifPaf \cite{kreiss2021openpifpaf} library is utilized. This allows each tracked pedestrian to be assigned a unique identifier that will be used to aggregate their total scores over time. The role of this trust dynamics is to introduce robustness in the estimation and to cater to additional constraints that might be imposed by the control system utilizing the trust estimation signal. For example, setting $\alpha + \beta = 1$ and $\beta_0 = 1$ in (\ref{agg_trust}) will model the dynamics as a moving average to smooth out disturbances in the estimation of total score. Moreover, setting $\alpha = 1$ and $\beta, \beta_0 \ll 1$ will model trust as monotonically increasing over time with saturation at 1. This is particularly useful for control applications where decreasing trust for a perceived agent can shrink the safe set, which might unintentionally remove the current state from the safe set.  More details regarding the effect of parameterization with an example comparison will be presented in a later part in this section.

Three specific behavioral indicators of trust are emphasized in this paper, i.e., $N_s = 3$, which are as follows:
\begin{itemize}
    \item smartphone usage $s_{1,j}$
    \item eye contact $s_{2,j}$
    \item pose fluctuation $s_{3,j}$
\end{itemize}
These traits, while not exhaustive, play a significant role in estimating pedestrian trust. The following subsections will elaborate on the quantification methodology for each trait.

\subsection{Smartphone Engagement Detection}

\begin{figure}
  \centering
  \includegraphics[width=0.8\columnwidth]{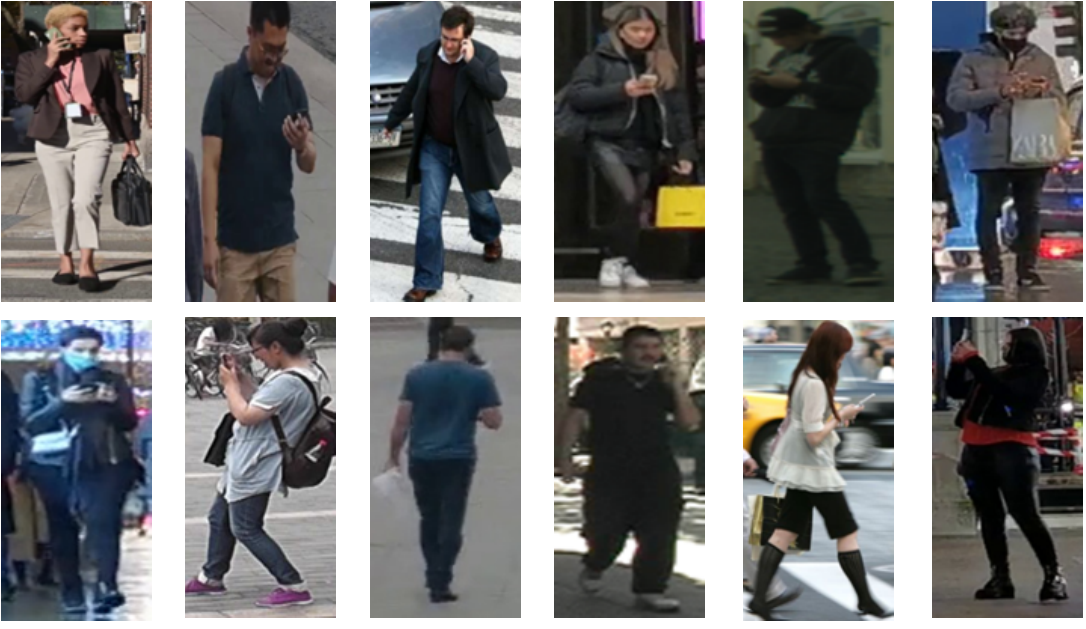}
  \caption{Example images of pedestrians engaged with their smartphones in a variety of poses, backgrounds, and lighting conditions}
  \label{smato_true}
\end{figure}

This paper contributes a curated dataset and a transfer-learning based model to classify smartphone usage, which are publicly available\footnote{https://github.com/saadejazz/smato}. The dataset consisting of 13,866 pedestrian images (3,770 with smartphone engagement) is compiled from images from various publicly available pedestrian datasets, such as PETA \cite{deng2014pedestrian}, PRW \cite{zheng2017person}, Penn-Fudan \cite{wang2007object}, and Cityscapes \cite{cordts2015cityscapes}, and from open-source images and videos from the internet. The images present diverse lighting conditions and camera angles, while covering a wide range of smartphone engagement scenarios, including eye engagement, where pedestrians are visually engaged with their smartphones, as well as phone call distractions. This variability in smartphone engagement scenarios can be seen in the example images shown in Fig.~\ref{smato_true}.

\begin{figure}
  \centering
\includegraphics[width=1\columnwidth]{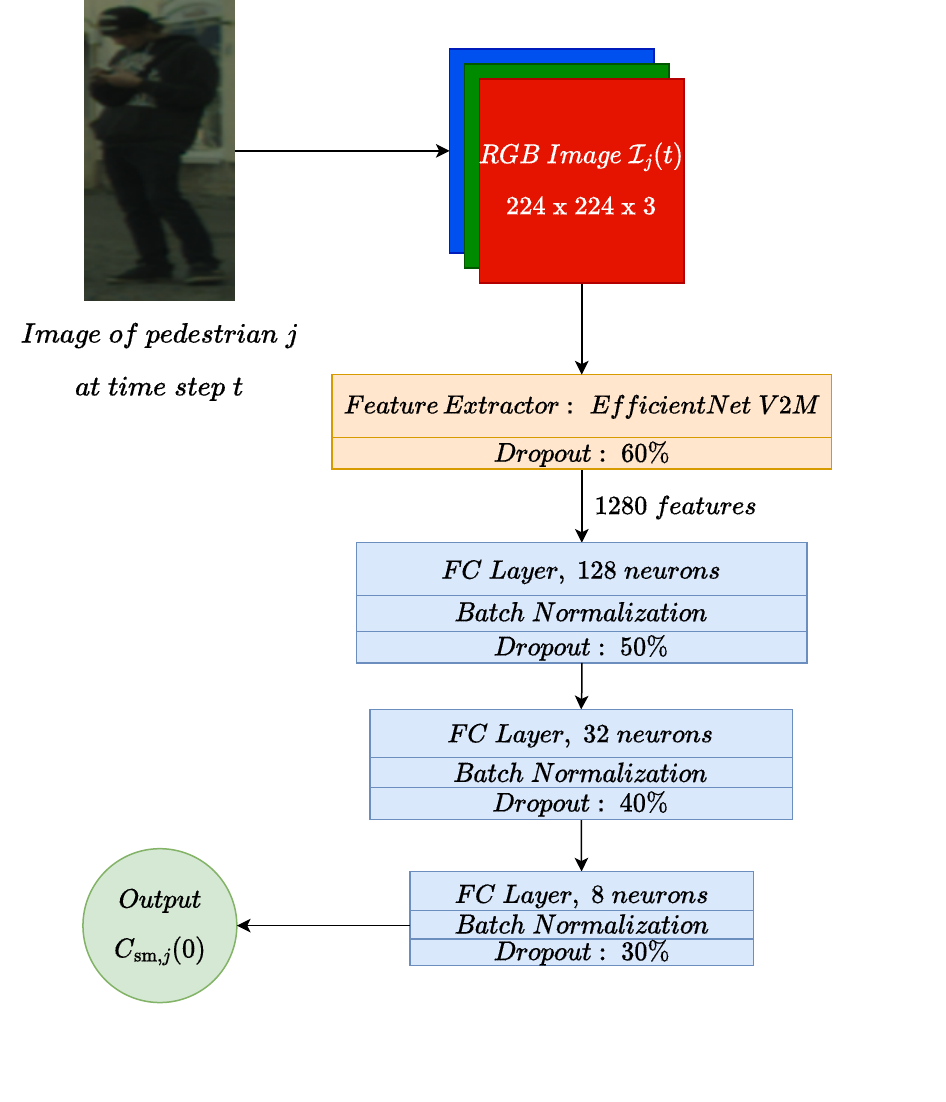}
  \caption{Model for Smartphone Usage Classification - EfficientNet V2 feature extractor followed by a three-layered fully-connected classification head}
  \label{network_smato}
\end{figure}
A transfer learning model was used to leverage pre-trained deep learning models. The binary classifier architecture can be visualized in Fig.~\ref{network_smato}, which consists of a feature extractor (EfficientNet V2 \cite{tan2021efficientnetv2}), followed by a fully connected classification head. The output of the network is a continuous value as confidence for presence of smartphone engagement. Continuous values are used to account for situations where definitive classifications are not possible. The trained model achieved an F1 score of 87.72\% and an accuracy of 92.97\% on the test dataset, demonstrating accurate detection of smartphone engagement in pedestrian images.

To incorporate temporal information, an aggregation scheme based on moving average is introduced. Trait scores obtained from a series of images of the same tracked pedestrian over time are aggregated to add robustness to the estimation. The following formulation considers the trait score $s_{1,j}$ for smartphone usage in pedestrian $j$:
\begin{equation}
    s_{1,j}(t) = \nu_1s_{1,j}(t-1) + (1-\nu_1)(1 - {C}_{{\rm sm},j}(t)),
\label{smato_score}
\end{equation}
where $\nu_1\in (\,0,1\,]$ is a positive constant, ${C}_{{\rm sm},j}$ represents the the confidence value for smartphone usage, and the initial value of $s_{1,j}(t)$ is given by 
\begin{equation*}
    s_{1,j}(0) = \nu_{01}(1 - {C}_{{\rm sm},j}(0)),
\end{equation*}
for some positive constant $\nu_{01} \in (\,0,1\,]$.

The confidence value ${C}_{{\rm sm},j}(t)$ is generated by a neural network model taking an isolated pedestrian image $\mathcal{I}_{j}(t)$ at time step $t$ as input, namely
\begin{equation*}
{C}_{{\rm sm},j}(t) = {\rm NN}_{\rm sm}(\mathcal{I}_{j}(t)).
\end{equation*}
During training, images with smartphone engagement were labeled as 1, so the confidence value is inversely related to the trait score. For continuity and robustness in estimation, a proportion $\nu_{1}$ of the previous trait score is maintained at each time step. Additionally, $\nu_{01}$ represents the proportion of the neural network output when the pedestrian was initially detected, serving as the initial trust score. By setting $\nu_{01}$ to 1, the initial confidence value contributes fully to the trait score.

\subsection{Eye Contact Detection}

\begin{figure}
  \centering
  \includegraphics[width=0.5\columnwidth]{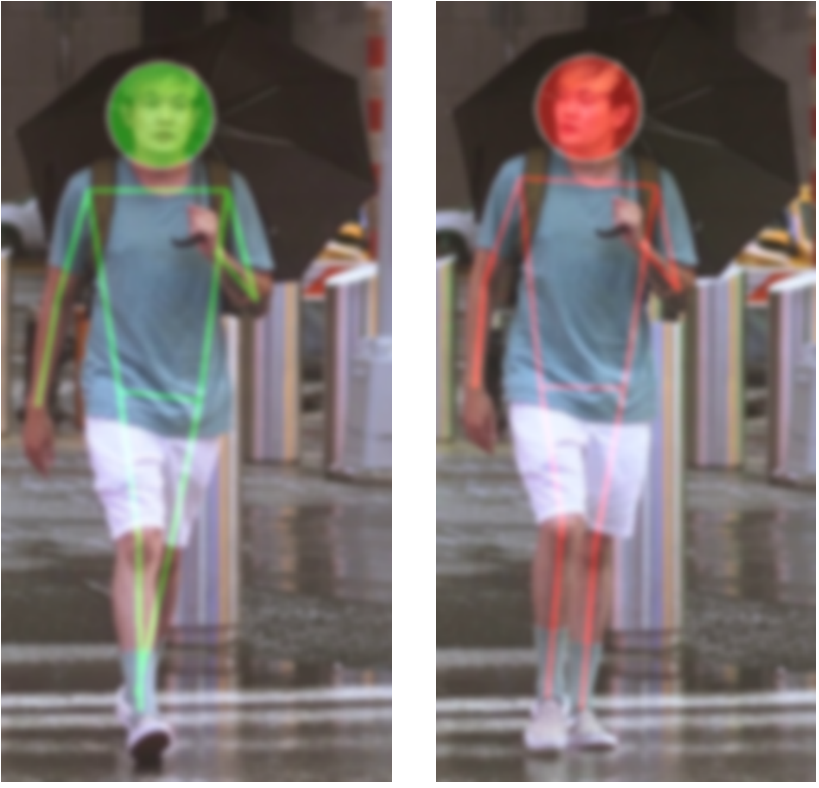}
  \caption{Eye contact detection using pose estimation \cite{belkada2021pedestrians} - green pose skeleton on the left image indicates presence of eye contact while the red pose skeleton on the right image indicates otherwise}
  \label{eye_contact}
\end{figure}

Eye contact detection is vital for estimating pedestrian trust and attentiveness in autonomous driving systems, as it provides insights into their awareness. The work of \cite{belkada2021pedestrians} on eye contact detection is incorporated as a behavior trait in the proposed trust estimation framework. The approach uses pose keypoints of pedestrians to determine the presence of eye contact, using a neural network. This pre-trained neural network is used in our work to output confidence values for the presence of eye contact. Fig.~\ref{eye_contact} visually illustrates this concept with a cropped image of the same pedestrian at different time instances. Green keypoints indicate a high probability of eye contact when the pedestrian faces the camera, while red keypoints indicate a low probability when the pedestrian looks away. 

\begin{figure*}
  \centering
  \includegraphics[width=1.95\columnwidth]{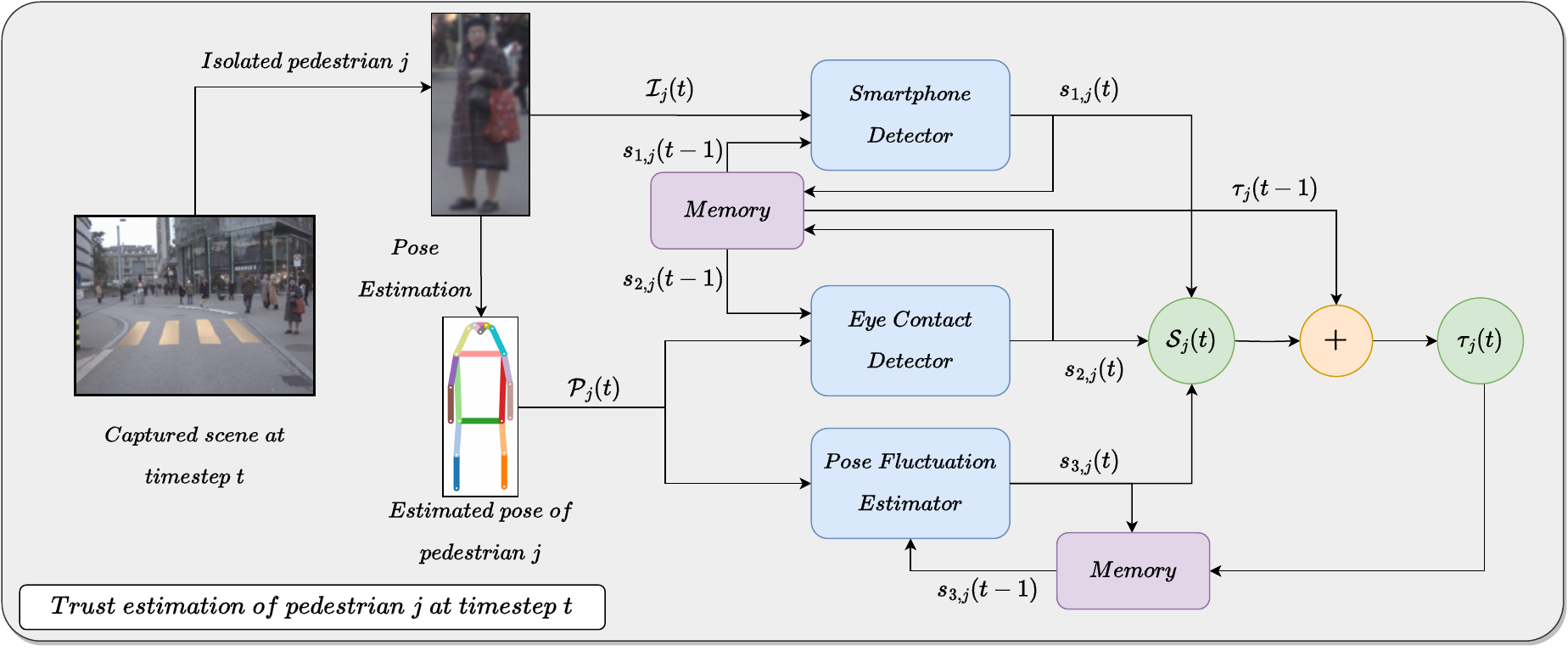}
  \caption{The complete trust estimation from captured RGB images of pedestrians - Memory in analogous to history i.e. the estimator stores the trait scores, trust values, and aggregated trust for at least one timestep in the past}
  \label{combined}
\end{figure*}

The aggregation scheme for the eye contact behavior trait involves iterative updates to the trait score by adding a small proportion of the new confidence value, resulting in a monotonically increasing trait score. Since humans rely on memory to keep track of surrounding vehicles, continuous eye contact is not maintained during an interaction. Therefore, trust gained from previous instances of established eye contact should persist as long as the pedestrian is tracked. This can be formulated as follows, with the eye contact trait score $s_{2, j}$ for pedestrian $j$:
\begin{equation}
    s_{2,j}(t) = \mbox{sat}_{[0,1]}(s_{2,j}(t-1) + \nu_2 {C}_{{\rm eye},j}(t)),
\label{eye_score}
\end{equation}
where $\nu_2\in (\,0,1\,]$ is a positive constant that is sufficiently small to avoid premature saturation of the trait score. The initial trait score $s_{2, j}(t)$ is given by
\begin{equation*}
    s_{2,j}(0) = \nu_{02}{\rm C}_{\rm eye}(0),
\end{equation*}
for some positive constant $\nu_{02}\in (\,0,1\,]$, which represents the proportion of initial confidence to be considered  for the initial trait score.

The confidence value ${C}_{{\rm eye},j}$ is generated by the output of the aforementioned trained neural network model that takes as input the pose keypoints $\mathcal{P}_j(t)$ of pedestrian $j$ at time step $t$ as input, namely
\begin{equation*}
{C}_{{\rm eye},j}(t) = {\rm NN}_{\rm eye}(\mathcal{P}_{j}(t)). \nonumber
\end{equation*}
The pose $\mathcal{P}_j$ comprises the positional coordinates of $N_k$ body keypoints in the image. In particular, we let $N_k = 17$ in this paper.

\subsection{Pose Fluctuation Computation}
Body pose dynamics offer valuable insights into a pedestrian's attentiveness and engagement with their surroundings. Pose fluctuation refers to the variations and movements in a pedestrian's body posture over time. By tracking and analyzing these changes, it becomes possible to infer their level of distraction or engagement. A steady and consistent body pose indicates higher attentiveness, while frequent or irregular pose changes suggest distraction or lack of focus.

Estimating pose fluctuation is relatively straightforward since pose is already estimated at each timestep for the purposes of eye contact detection. The estimation of this trait score involves calculating the deviation of body keypoints relative to the pedestrian's bounding box. 
Let $P_{j,t}$ be the pose keypoints relative to the bounding box of pedestrian $j$ at timestep $t$, with $P_j(t)[n]$ and $\mathcal{P}_j(t)[n]$ corresponding to the position of the $n^{th}$ body keypoint in relative and absolute terms respectively. The relative keypoints $P$ are determined from the absolute pose keypoints $\mathcal{P}$ as follows (the indices $j$ and $t$ are omitted for simplicity):
\begin{equation}
    P[n] = (\mathcal{P}[n] - X_{bbox}) \;\oslash\; D_{bbox},
\end{equation}
where $X_{bbox}$ is the coordinates of the top left corner of the pedestrian bounding box, $D_{bbox}$ is the dimensions of the bounding box (width and height), and $\oslash$ is the element-wise division operator.

The confidence value associated with pose fluctuation at timestep $t$, ${C}_{{\rm fluc},j}$ is then given by
\begin{equation}
    {C}_{{\rm fluc},j}(t) = \mbox{sat}_{[0,1]}\left(\frac{\mathcal{F}N_{k}}{\sum_{n=1}^{N_k}\norm{P_j(t)[n] - P_j(t-1)[n]}_2}\right),
\label{pose_conf}
\end{equation}
where $\mathcal{F}$ is the \textit{fluctuation sensitivity} which is the constant of proportionality between the confidence value, outputted from ${C}_{{\rm fluc},j}$, and the mean deviation in relative pose (they are inversely proportional), and $\norm{\cdot}_2$ is the L2 norm. Here, recall $N_k=17$, which is the total number of body keypoints in the estimated pose.
The trait score associated with pose fluctuation, $s_{3, j}$ is determined as follows:
\begin{equation}
    \begin{gathered}
    s_{3,j}(t) = \nu_3s_{3,j}(t-1) + (1-\nu_3){C}_{{\rm fluc},j}(t),
    \end{gathered}
\label{pose_score}
\end{equation}
where $\nu_1\in (\,0,1\,]$ is a positive constant, and the initial value of $s_{3, j}(t)$ is given by
\begin{equation*}
    s_{3,j}(0) = \nu_{03},
\end{equation*}
for some positive constant $\nu_{03} \in (\,0,1\,]$ that represents the initial trait score.

\subsection{Overall Structure of Trust Estimator}
The overall structure of the trust estimator can be seen in Fig.~\ref{combined}. All three trait scores $s_{1, j}(t)$, $s_{2, j}(t)$, and $s_{3, j}(t)$ corresponding to smartphone usage, eye contact, and pose fluctuation of pedestrian $j$ at timestep $t$ are determined using the image captured from the camera of the ego agent at that timestep. This image goes through a pedestrian isolation step, where all pedestrians are isolated and uniquely identified (for tracking purposes). The isolated image of pedestrian $j$ at timestep $t$, $\mathcal{I}_j(t)$ is used for two purposes: to compute $s_{1, j}(t)$ and to determine pose $\mathcal{P}_j(t)$. The determined pose is used to compute the remaining trait scores $s_{2, j}(t)$ and $s_{3, j}(t)$. These trait scores are linearly combined as formulated in (\ref{lin_comb}) to output the total score for the pedestrian $\mathcal{S}_j(t)$. This combined with the pedestrian's prior trust $\tau_j(t-1)$ according to the formulation presented in (\ref{agg_trust}) results in their new trust value for timestep $t$.

\begin{figure}
  \centering
  \includegraphics[width=0.75\columnwidth]{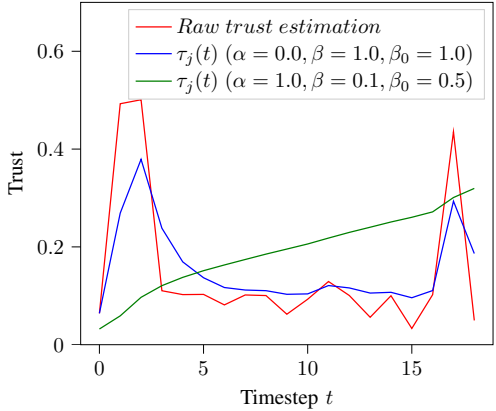}
  \caption{Comparison of different trust dynamics - with raw estimates of trust based on confidence scores only}
  \label{dynamic}
\end{figure}

It is important to note that the system keeps track of trait scores and trust value for one timestep in the past for use in the formulations presented in this section, that improve robustness of the estimations. We can understand the importance of the involved aggregation schemes be referring to the plot in Fig.~\ref{dynamic} and (\ref{agg_trust}), which compares the two types of smoothing discussed earlier with a trust signal only dependent on the current timestep. This raw estimation can be seen in the plot as a red line and is a result of setting the recursive variables ($\alpha$, $\nu_1$, $\nu_2$, and $\nu_3$) to zero. The trust estimate shown in blue is a smoothing variant with $\nu_1, \nu_2, \nu_3 > 0$. Even with $\alpha=0$, one can notice the robustness to minor disturbances in the raw trust estimation signal, especially when $4 < t < 16$. Increasing the value of $\alpha$ can introduce further smoothing which can be tuned according to preference. Another example can be seen as the green line in the plot which depicts how the trust dynamics can accommodate additional constraints, which in this case is the requirement of being monotonically increasing i.e., $\tau_j(t)\geq\tau_j(t-1)$. As stated earlier in this section, such conditions might be necessary to avoid shrinkage of the safe set of a control system that is based on the trust estimate. This can be done by setting $\alpha=1$ and $\beta$ as the rate of increase in trust per timestep, which can be tuned depending on the sampling rate of the trust estimation process.

\subsection{Examples of Trust Estimation}
\begin{figure}
  \centering
  \includegraphics[width=0.7\columnwidth]{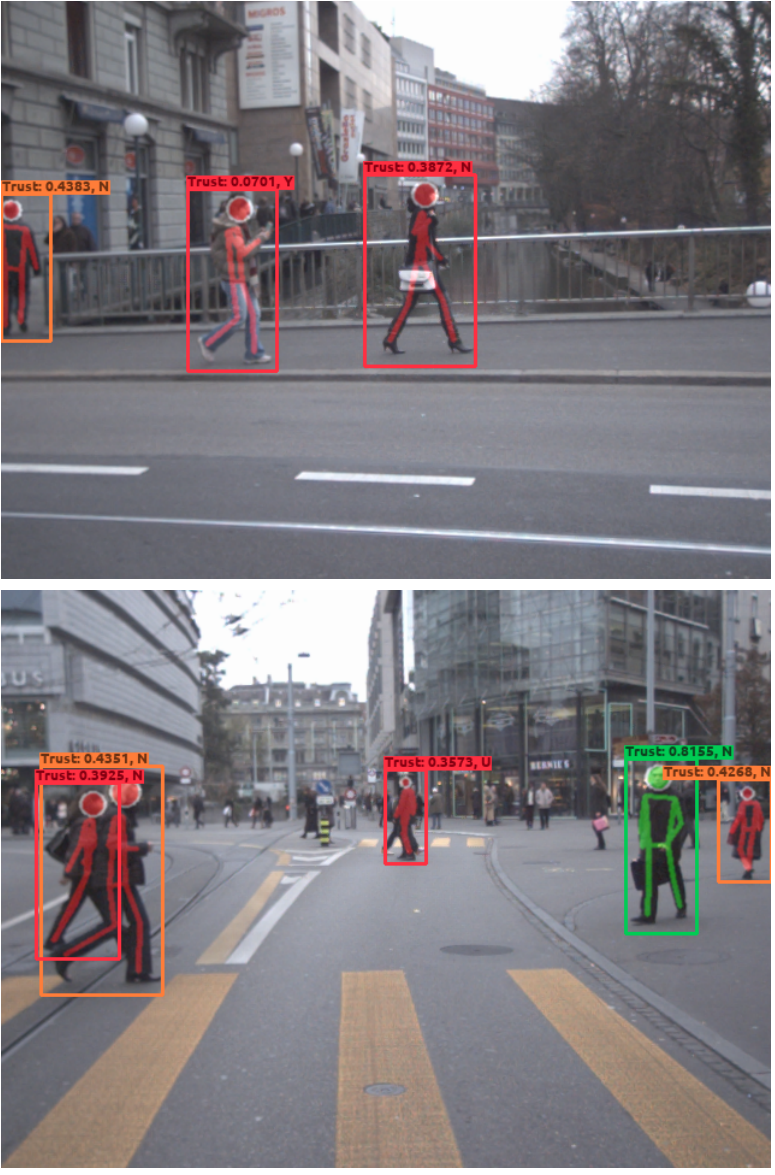}
  \caption{Trust estimation of pedestrians in roadside images - attentive pedestrians have a high trust score indicated by a green bounding box, distracted pedestrians are marked with a red bounding box, and an orange bounding box indicates a mediocre level of trust}
  \label{example1}
\end{figure}

Fig.~\ref{example1}  demonstrate the results of the trust estimation on an example image with multiple pedestrians. The color of the bounding boxes indicate the trust levels which are displayed additionally on the top of the box. Furthermore, the algorithm also produces decent results for images of simulated pedestrians (from the CARLA simulator, the results of which will be discussed in Section \ref{section-simulation}, specifically see Fig.~\ref{trust_frames}). This indicates that the classifiers in the trust estimator have learned meaningful representations of trust, with a focus on body posture and joint positions instead of unrelated correlations. Complete details of the visualization scheme along with additional examples can be viewed at the GitHub repository\footnote{https://github.com/saadejazz/trusty} for our trust estimator.

\section{Trust-aware Safe Navigation}
\label{section-control}
\subsection{Preliminary: Control Barrier Function}
In the following discussion, we let 
$x_{\rm e}$, $v_{\rm e}$, $x_j$, and $v_j$ denote the position and velocity of the ego vehicle and those of detected pedestrian $j \in J=\{1,2,\ldots, N_p\}$, respectively. 
For simplicity of notation, we further let 
$\mathbf{x}_{\rm e}=[x_{\rm e}, v_{\rm e}]$ and
$\mathbf{x}_{\rm p}=[x_1, x_2, \hdots, x_{N_p}, v_1, v_2, \hdots, v_{N_p}]$.

In safety-critical control, the objective is to keep the system within a designated safe set denoted by $C_{\rm safe}$. 
First, we consider the safety control of the ego vehicle to each pedestrian.
To this end, a continuously differentiable function $\phi({x}_{\rm e},x_j)$ is defined. A quadratic function, as in 
\begin{equation}
    \phi({x}_{\rm e},x_j) = \norm{x_{\rm e}-x_j}^2 - R^2,
\label{h_quad}
\end{equation}
is commonly used in the context of CBFs \cite{lyu2022responsibilityassociated, zeng2021mpccbf} for maintaining a safe distance, denoted by $R$ between the ego agent at position $x_{\rm e}$ and some pedestrian at position $x_j$. The term \textit{pedestrians} is used interchangeably with \textit{obstacles} and \textit{other agents} to promote generality.
The overall safe set is defined by using the vector-valued function $h(\mathbf{x}_{\rm e},\mathbf{x}_{\rm p})$
\begin{equation}
h(\mathbf{x}_{\rm e},\mathbf{x}_{\rm p})= \begin{bmatrix}
           \phi({x}_{\rm e},x_1)& \cdots & \phi({x}_{\rm e},x_{N_p})\\
         \end{bmatrix}
\end{equation}
as
\begin{equation}
C_{\rm safe}(\mathbf{x}_{\rm p}) = \left\{ \mathbf{x}_{\rm e} \in \mathcal{X}  : h(\mathbf{x}_{\rm e},\mathbf{x}_{\rm p}) \geq 0 \right\},
\end{equation}
where symbol $\geq$ represents the element-wise inequality. 
In $C_{\rm safe}(\mathbf{x}_{\rm p})$, the function $h(\mathbf{x}_{\rm e},\mathbf{x}_{\rm p})$ is 
 used as a barrier to prevent the system from entering unsafe regions \cite{ames2019control}.

For a function $h$ to be considered a set of CBFs, it must satisfy certain conditions. The partial derivative of $h(\mathbf{x}_{\rm e},\mathbf{x}_{\rm p})$ with respect to $\mathbf{x}_{\rm e}$ should not be zero for all $\mathbf{x}_{\rm e}$ on the boundary of the safe set $\partial C_{\rm safe}(\mathbf{x}_{\rm p})$. This ensures that $h(\mathbf{x}_{\rm e})$ captures the proximity to the boundary. Additionally, there should exist an extended class $\mathcal{K}_{\infty}$ vector-valued function $\kappa$ and a valid control action such that the time derivative of $h(\mathbf{x}_{\rm e})$ along the system trajectory satisfies (\ref{cbf_cont}), ensuring that the system remains within the safe set.
\begin{equation}
    \dot{h}(\mathbf{x}_{\rm e}, \mathbf{x}_{\rm p}) \geq -\kappa(h(\mathbf{x}_{\rm e}, \mathbf{x}_{\rm p}))   
\label{cbf_cont}
\end{equation}
The condition (\ref{cbf_cont}) can also be extended to discrete-time CBFs \cite{zeng2021mpccbf} at each time step, as shown in 
\begin{equation}
    \Delta h(\mathbf{x}_{\rm e}(t), \mathbf{x}_{\rm p}(t)) \geq -\gamma(t) \odot h(\mathbf{x}_{\rm e}(t), \mathbf{x}_{\rm p}(t))
\label{cbf_discrete}
\end{equation}
where symbol $\odot$ represents the Hadamard product, 
 $\Delta h(\mathbf{x}_{\rm e}(t), \mathbf{x}_{\rm p}(t))$ is a vector-valued function given by
\begin{equation*}
    \Delta h(\mathbf{x}_{\rm e}(t), \mathbf{x}_{\rm p}(t)) := h(\mathbf{x}_{\rm e}(t+1), \mathbf{x}_{\rm p}(t+1)) - h(\mathbf{x}_{\rm e}(t), \mathbf{x}_{\rm p}(t))
\end{equation*}
and $\gamma(t) = [\gamma_1(t), \ldots, \gamma_{N_p}(t)]\in [\,0,1\,]^{N_p}\subset \mathbb{R}^{N_p}$ determines the control aggressiveness at time $t$ i.e., a lower gamma would enforce a more conservative approach and vice versa.

\subsection{Trust-adaptive CBF constraints}

This work adopts discrete-time CBF constraints (\ref{cbf_discrete}) where the parameter $\gamma$ governs the safety margin and depends on the estimated trust, in other words, $\gamma(t) = \gamma(\tau(t))$.

Since $\gamma$ influences the aggressiveness of control, it provides an intuitive inclusion of the trust signal estimated from the trust estimator shown in Fig.~\ref{combined} and computed as in (\ref{agg_trust}). The adaptive update of the parameter $\gamma_j(t)$, which is the $j$th element of $\gamma(t)$, for pairwise CBF distance-based constraints between the ego agent and all pedestrians $j \in J$ is determined by incorporating the trust signal of each pedestrian $\tau_j(t)$ at time $t$. This dynamic update process ensures that the parameter $\gamma_j(t)$ is influenced by the trust level of the corresponding pedestrian at each time instance. The adaptive discrete-time CBF can be reformulated for pedestrian $j$ at timestep $t$ as in (\ref{cbf_discrete}) with the following parameter update
\begin{equation}
    \gamma_j(t) =  \gamma_{\rm ini} + \delta \tau_j(t)^{\lambda},
\label{weight}
\end{equation}
where $\gamma_{\rm ini}$ is the positive constant representing the base value when trust is either zero or unavailable; $\delta$ is the positive constant representing the trust sensitivity, which determines how a unit change in trust affects $\gamma_j$ and hence control aggressiveness; and $\lambda$ is the constant representing trust penalty, which governs the penalty per unit decrease in trust and is important for relative safety considerations i.e. when navigating through multiple pedestrians. Effectively, these hyperparameters allow control over the shape of the function $\gamma_j$ with respect to the trust signal $\tau_j$ providing a way to regulate the impact of trust in the control policy. Ideally, this tuning needs to be part of the design
process of the control system to accommodate intended control preference and/or style.

Furthermore, to ensure conformance with the constraint $\gamma_j(t) \in [\,0,1\,]$, the hyperparameters must satisfy the following conditions:
\begin{subequations}
    \begin{align}
    \gamma_{\rm ini} + \delta \leq 1\\
    \lambda \geq 1,
    \end{align}
\end{subequations}
provided which the discrete-time CBF in (\ref{cbf_discrete}) can be used to guarantee safety.

\subsection{Proposed Control Framework}
\begin{figure*}
  \centering
  \includegraphics[width=1.1\columnwidth]{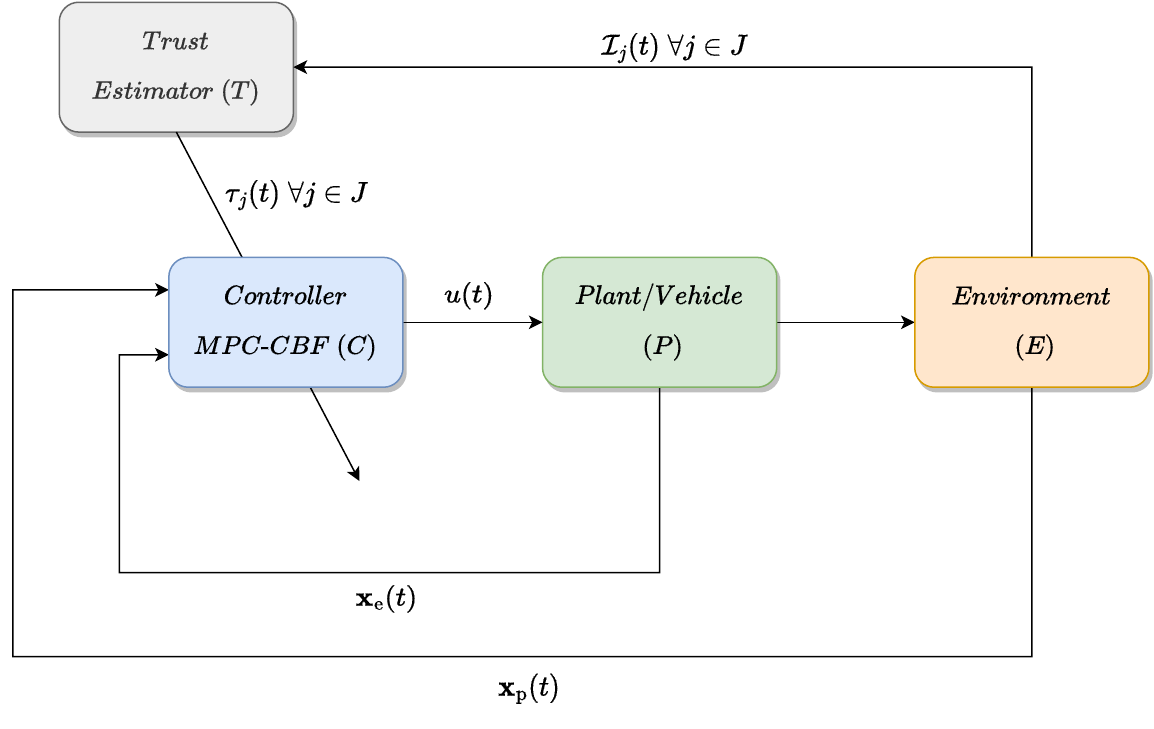}
  \caption{Block diagram of the proposed trust-adaptive MPC-CBF framework. At timestep $t$, the controller $C$ receives feedback from the plant $P$ and environment $E$, along with trust estimates of all pedestrians from the trust estimator (Fig.~\ref{combined}) as parameter updates to the CBF constraints in the controller. The optimization problem is solved for a finite future horizon, with the current control input $u(t)$ applied to $P$}
  \label{control-frame} 
\end{figure*}
This work incorporates a barrier function formulation as in (\ref{h_quad}) and (\ref{cbf_discrete}) with a trust-adaptive $\gamma$ value introduced in (\ref{weight}), to impose respective safety considerations based on trust estimates from the trust estimator, as shown in Fig.~\ref{combined}, in an MPC framework. The CBF constraints are set for a finite time horizon in the future which adds further robustness to the system against dynamic agents in the environment. Not only does this ensure that the safety margin $R$ is respected in the future, but also that the trust-dependent safety considerations are maintained. This can be observed in the numerical simulations that will be presented later in this section.

For the MPC formulation, consider the discrete-time prediction model that describes the evolution of the state of the ego vehicle $\mathbf{x}_{\rm e}$ based on control input $u$:
\begin{equation}
    \hat{\mathbf{x}}_{\rm e}(t+1) = f_{\rm e}(\mathbf{x}_{\rm e}(t), u(t)),
\label{f_e}
\end{equation}
where $\hat{\mathbf{x}}_{\rm e}(t+1)$ is the predicted model state at timestep $t+1$.

The optimization problem is formulated to minimize a cost function with respect to the sequence of control inputs $\mathbf{u}(t:t+N_h-1) = [\,u(t), u(t+1), \hdots, u(t+ N_h -1)\,]$, where $N_h$ is the finite prediction horizon, as follows
\begin{subequations}
\begin{align}
& \min_{\mathbf{u}(t:t+N_h-1)}  p(\hat{\mathbf{x}}_{\rm e}(t+N_h)) + \sum_{k=0}^{N_h-1} q(\hat{\mathbf{x}}_{\rm e}(t+k), u(t+k)) \label{cost}\\
&\text{s.t.} \quad \nonumber \\
& \hat{\mathbf{x}}_{\rm e}(t+k+1) = f_{\rm e}(\hat{\mathbf{x}}_{\rm e}(t+k), u(t+k)), \;\; \forall \; k \in K, \label{pred}\\
& \hat{\mathbf{x}}_{\rm e}(t+k) \in \mathcal{X}, \; u(t+k) \in \mathcal{U}, \;\; \forall \; k  \in K, \label{bounds}\\
& \hat{\mathbf{x}}_{\rm e}(t + N_h) \in \mathcal{X}, \label{bound}\\
 & \hat{\mathbf{x}}_{\rm e}(t) = \mathbf{x}_{\rm e}(t), \; \hat{\mathbf{x}}_{\rm p}(t) = \mathbf{x}_{\rm p}(t),\\
& \hat{\mathbf{x}}_{\rm p}(t + k + 1) = f_{\rm p}(\hat{\mathbf{x}}_{\rm p}(t + k)), \;\; \forall \; k \in K,\label{ped_predict}\\
&  h(\hat{\mathbf{x}}_{\rm e}(t+k+1), \hat{\mathbf{x}}_{\rm p}(t+k+1)) \nonumber \\ 
& \hspace{.5cm} \geq ({\bf 1}_{N_p}-\gamma(t)) \odot h(\hat{\mathbf{x}}_{\rm e}(t+k), \hat{\mathbf{x}}_{\rm p}(t+k)),\ \forall \; k \in K, \label{cbf_imp}
\end{align}
\label{proposed}
\end{subequations}
where symbol ${\bf 1}_{N_p}$ represents the all-one vector of length $N_p$.

In this control formulation\footnote{It is important to note that in the control formulation in this section, while the time variable $t$ used is the same as the one used in the trust estimation process, it is entirely possible, even necessary, to use different sampling times for the trust perception and trust-based control systems.}, $K$ is the set of timesteps in the future, $K = \{0, 1, \hdots, N_h-1\}$ and $J$ is the set of indices of the pedestrians as introduced in (\ref{lin_comb}). Recall that $N_p$ is the number of pedestrians detected in the environment. The prediction model in (\ref{pred}) corresponds to (\ref{f_e}) and the state of the ego vehicle and the control inputs are bounded within the sets $\mathcal{X}$ and $\mathcal{U}$ respectively as in (\ref{bounds}) and (\ref{bound}). The functions $p$ and $q$ represent the terminal and stage costs respectively. At each time step $t$, the ego vehicle has the state measurements $\mathbf{x}_{\rm e}(t)$ and $\mathbf{x}_{\rm p}(t)$, and trust estimates $\tau_j(t)$ for all pedestrians $j \in J$. Moreover, a simplified prediction model for the position of pedestrians, given in (\ref{ped_predict}), is utilized as follows:
\begin{align}
    f_{\rm p}(\mathbf{x}_{\rm p}(t + k)) &=\begin{bmatrix}
           x_1(t+k) + v_{1}(t)\Delta t\\
           \vdots \\
           x_{N_p}(t+k) + v_{N_p}(t)\Delta t\\
           v_{1}(t)\\
           \vdots \\
           v_{N_p}(t)
         \end{bmatrix}^\top,
\end{align}


where $v_{j}(t)$, which is the velocity of pedestrian $j$ at timestep $t$, stays constant regardless of the prediction horizon. It is assumed that the ego vehicle will use a combination of sensors or image based processing to approximate this velocity, a process not part of this work. Moreover, $\Delta t$ is the step time of the designed control system.

Furthermore, (\ref{cbf_imp}) represents the set of $N_p \times N_h$ discrete-time trust-adaptive CBF constraints as in (\ref{cbf_discrete}) which is the essence of this proposed formulation to ensure safety considerations dependent on trust. The complete framework can be visualized in Fig.~\ref{control-frame}.

\subsection{Numerical Simulations}

To validate the proposed control design (\ref{proposed}), the optimization the problem is solved using using SLSQP \cite{kraft1988software} for simplistic scenarios in a 2D 50x50 unit squared grid with stationary or moving pedestrians (referred to as obstacles in this subsection). The ego agent's model is a single integrator formulated in as a discrete-time model (\ref{f_e}) with
\begin{equation}
    f_e(\mathbf{x}_e(t), u(t)) = \mathbf{x}_e(t) + u(t) \Delta t
\end{equation} 
The cost function aims to track a reference input velocity generated by a bounded proportional controller. The reference generator takes the current state and goal state as inputs to produce the desired velocity $\bar{u}$ for each timestep in the prediction horizon. For the cost function of the optimization problem, we let $q$ of (\ref{cost}) as:
\begin{equation}
    q(\hat{\mathbf{x}}_{\rm e}, u) = \|u - \bar{u}\|^2
\end{equation}
Furthermore, the following variables are kept constant throughout the simulations, unless otherwise specified: $\Delta t=0.05$, $R=3$, $\gamma_{\rm ini}=0.03$, $\delta=0.08$, and $\lambda=1.5$. For the sake of simplicity, $N_p \leq 2$, the trust and velocity of perceived agents is kept constant throughout each simulation scenario ($\tau_j(t) = \tau_j$ and $v_0(t) = v_0$), and the optimal control input determined at each timestep is realized by the autonomous agent without any latency.

  \begin{figure}
  \centering
  \includegraphics[width=0.75\columnwidth]{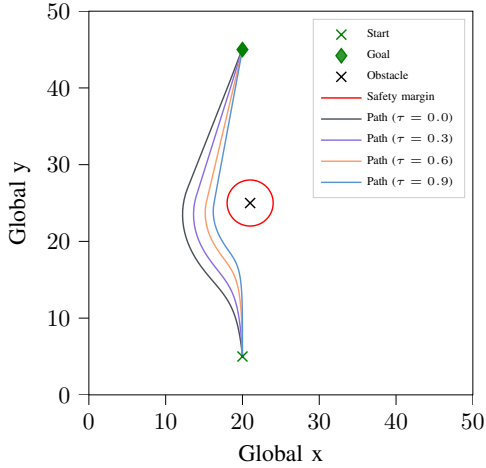}
  \caption{Scenario I: Effect of trust on safety distance - a lower trust value encourages a greater safety distance from the agent/obstacle.}
  \label{scene1}
\end{figure}

\textbf{Scenario I:} Effect of trust on safety margin for navigating around a single stationary agent. 

\textit{Initial Conditions: } $N_h=7$, $x_e(0) = [20, 5]$, $N_p = 1$, $x_1(0) = [21, 25]$, and $v_1 = [0, 0]$, while the goal location of the ego agent is $x_g = [20, 45]$, i.e. we aim $x_e(t) \rightarrow x_g$.

Several trust values in the range of $[\,0, 1\,]$ for the blocking agent are used to solve the optimization problem to understand the effect of trust. The results can be seen in Fig.~\ref{scene1}. The lower the value of trust, the farther the ego agent's path is from the safety margin of the agent whose trust is estimated. This is in line with the objective of this paper.

\begin{figure}
  \centering
  \includegraphics[width=0.75\columnwidth]{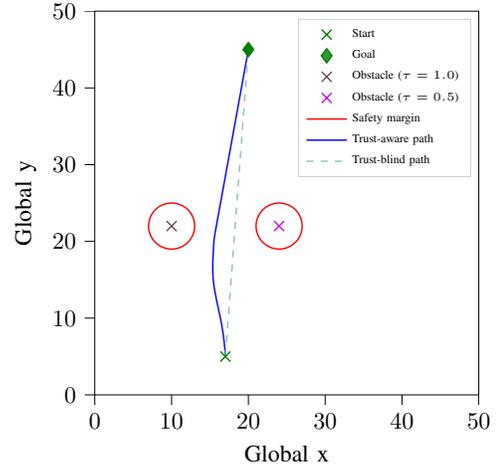}
  \caption{Scenario II: Navigation between agents with different trust levels - the ego agent stays closer to the high-trust agent than the low-trust agent}
  \label{scene2_1}
\end{figure}

\begin{figure*}
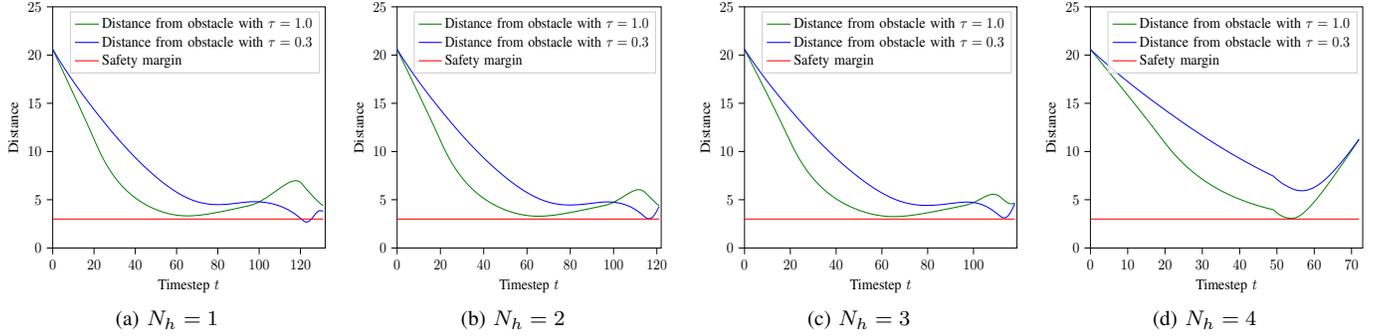

    \centering
    \subfloat[][$N_h=1$]{\includegraphics[width = 0.49\columnwidth, height=4cm]{plots/scene3_1.tex}\label{scene3_1}}
    \hfill
    \subfloat[][$N_h=2$]{\includegraphics[width = 0.49\columnwidth, height=4cm]{plots/scene3_2.tex}\label{scene3_2}}
    \hfill
    \subfloat[][$N_h=3$]{\includegraphics[width = 0.49\columnwidth, height=4cm]{plots/scene3_3.tex}\label{scene3_3}}
    \hfill
    \subfloat[][$N_h=4$]{\includegraphics[width = 0.49\columnwidth, height=4cm]{plots/scene3_4.tex}\label{scene3_4}}
    \caption{Scenario III: Effect of prediction horizon of MPC on maintaining adequate safety margin from other agents/obstacles. $N_h=1$ causes the ego agent to ignore dynamic changes in the environment resulting in a breach of safety margin. $N_h \in \{2, 3\}$ loses the benefits of trust-based navigation, even though safety margin is respected. $N_h=4$ is a sufficient prediction horizon considering this specific system setup, since trust-aware safety considerations are respected throughout the path. One can also notice that the time taken to navigate is less for $N_h=4$ compared to other values, which can be worth investigating further.}\label{pred_hor}
  \end{figure*}

\textbf{Scenario II:} Navigation through two stationary agents with a noticeable difference in trust. This simulates the path of a vehicle through a pedestrian crossing with two pedestrians on opposite sides of the crossing, with one of them visibly distracted. Ideally, the vehicle should maintain a larger safety margin from the inattentive pedestrian. 
 
 \textit{Initial Conditions: } $N_h=7$, $x_e(0) = [17, 5]$, $N_p = 2$, $x_1(0) = [10, 22]$, $x_2(0) = [24, 22]$, $v_1 = [0, 0]$, $v_2 = [0, 0]$, $\tau_1 = 1.0$, and $\tau_2 = 0.5$, while the goal location of the ego agent is $x_g = [20, 45]$.

Fig.~\ref{scene2_1} compares the path of the proposed trust-aware system when compared with the direct line path which was provided by the reference generator. The path of the trust-aware system keeps a larger distance from the less trustworthy agent (trust value was half that of the trustworthy agent which had the maximum value possible for trust), as expected. Moreover, it is important to note that changing the hyperparameters of the weight function in (\ref{weight}) will modify the path that the ego agent takes to improve safety, essentially providing the capability to tune the impact of trust in the control policy.

\textbf{Scenario III:} Navigation through two moving agents with different trust levels. This is to simulate the importance of the MPC that predicts the position of the agents to maintain appropriate safety margin based on trust, over a future horizon. 
 
\textit{Initial Conditions: } $x_e(0) = [20, 5]$, $N_p = 2$, $x_1(0) = [10, 23]$, $x_2(0) = [30, 23]$, $v_{1} = [2, 3]$, $v_2 = [-2, 3]$, $\tau_1 = 1.0$, and $\tau_2 = 0.3$, while the goal location of the ego agent $x_g = [20, 45]$. The obstacles also move in the direction of the goal to imitate adversarial conditions. 

The prediction horizon $N_h$ is varied from 1 to 4 to understand the significance of the MPC in the proposed approach. The distance plots shown in Fig.~\ref{pred_hor} demonstrate the effect of respective prediction horizon on the distance maintained by the ego agent from both the high-trust obstacle (shown in green) and the low-trust obstacle (shown in blue). Ideally, the green line should be below (or at the same level) the blue line to ensure that extra safety consideration is provided to the low-trust obstacle, at all timesteps. However, for lower values of prediction horizon, the ego agent fails to predict the future movement of the obstacle and can even violate the safety margin (red line), which is the case in Fig.~\ref{scene3_1} where $N_h=1$. For $N_h<3$, while the safety margin is respected, the effect of trust is ignored for $t>100$. Lastly, Fig.~\ref{scene3_4} demonstrates that $N_h=4$ is adequate to maintain safety margin while effectively employing the trust-based CBF constraints. This validates the importance of an adequate prediction horizon to compensate for dynamics of other agents in the system and to avoid reaching a state where violation of safety margin in the future becomes inevitable. It is important to tune the value of $N_h$ as part of the design process depending on the model complexity and unpredictability of other agents in the environment. A higher value of $N_h$ would enhance safety at the cost of computation time, and thus should be chosen accordingly. 

\section{Autonomous Driving Simulations}
\label{section-simulation}
\begin{figure}
  \centering
  \includegraphics[width = 0.8\columnwidth]{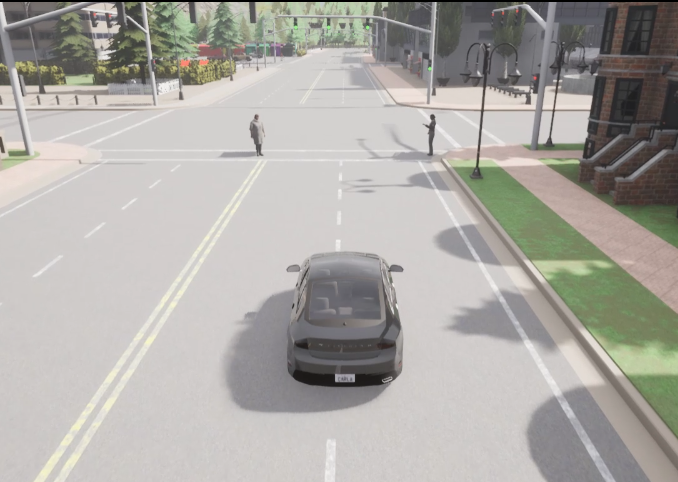}
  \caption{Simulation environment in CARLA}
  \label{carla_sim}
\end{figure}

\begin{figure}
  \centering
  \includegraphics[width = 0.5\columnwidth]{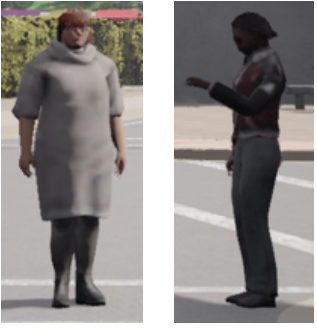}
  \caption{Types of simulated pedestrians - the high trust pedestrian on the left maintains a neutral stance facing the ego vehicle, and the low trust pedestrian on the right has bent neck and elbows, showing signs of smartphone engagement}
  \label{ped_type}
\end{figure}

\begin{figure*}
    \centering
    \subfloat[][Frame 1]{\includegraphics[width = 0.50\columnwidth, height=3cm]{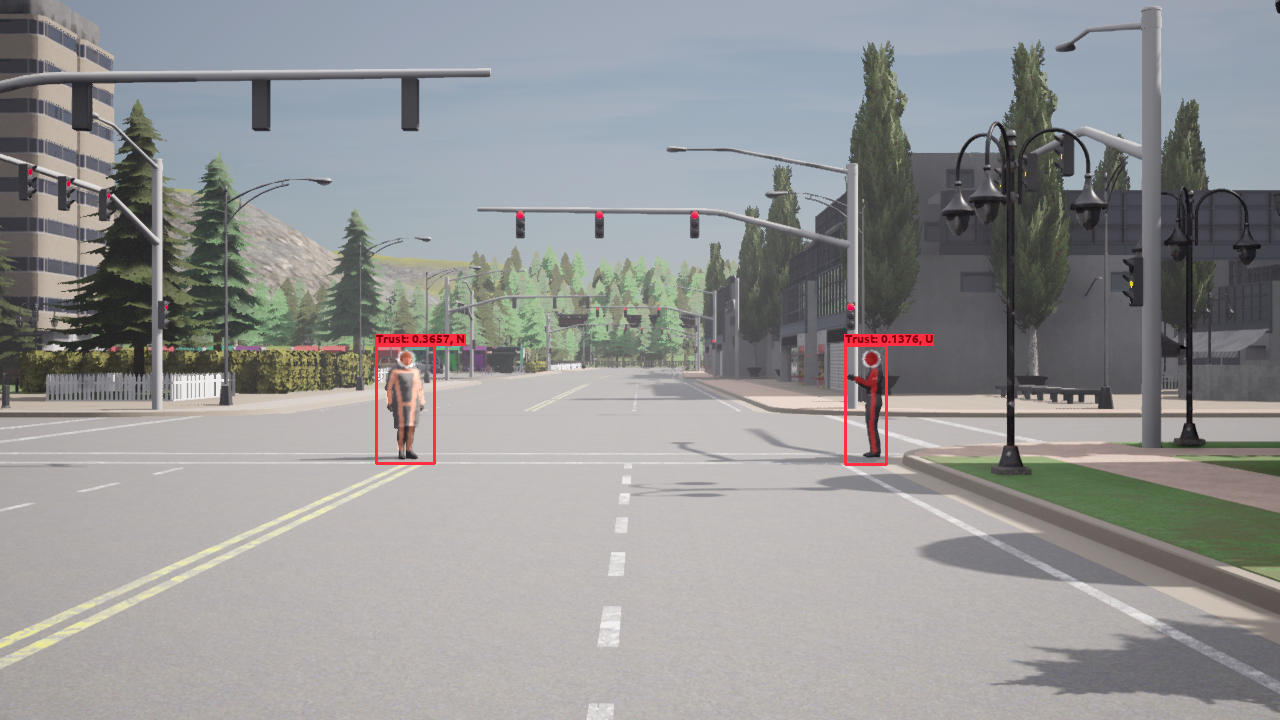}\label{frame1}}
    \hfill
    \subfloat[][Frame 4]{\includegraphics[width = 0.50\columnwidth, height=3cm]{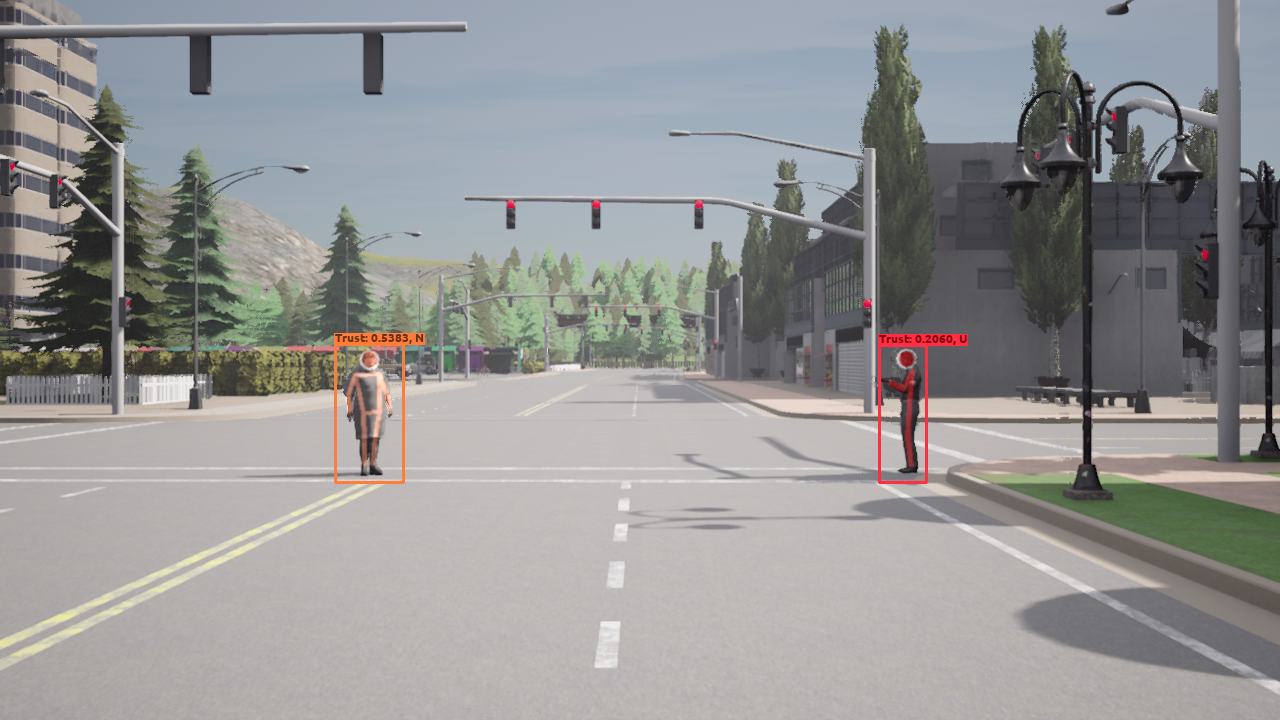}\label{frame2}}
    \hfill
    \subfloat[][Frame 7]{\includegraphics[width = 0.50\columnwidth, height=3cm]{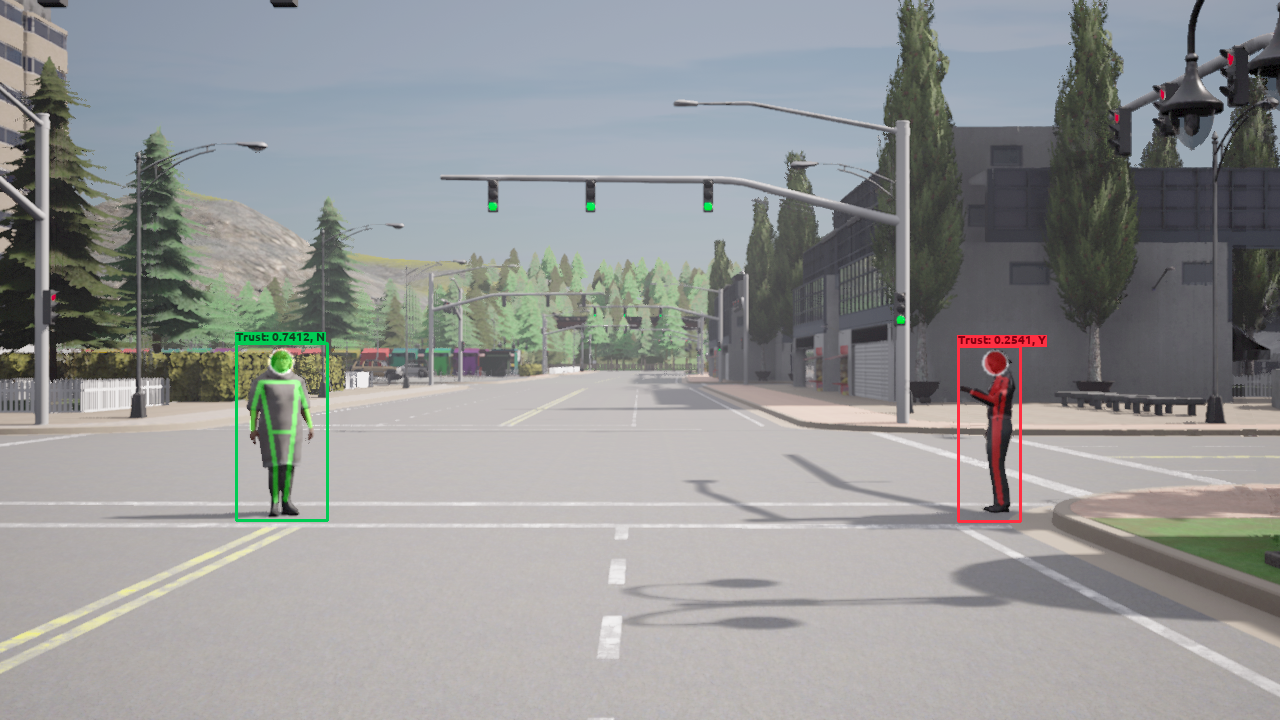}\label{frame3}}
    \hfill
    \subfloat[][Frame 10]{\includegraphics[width = 0.50\columnwidth, height=3cm]{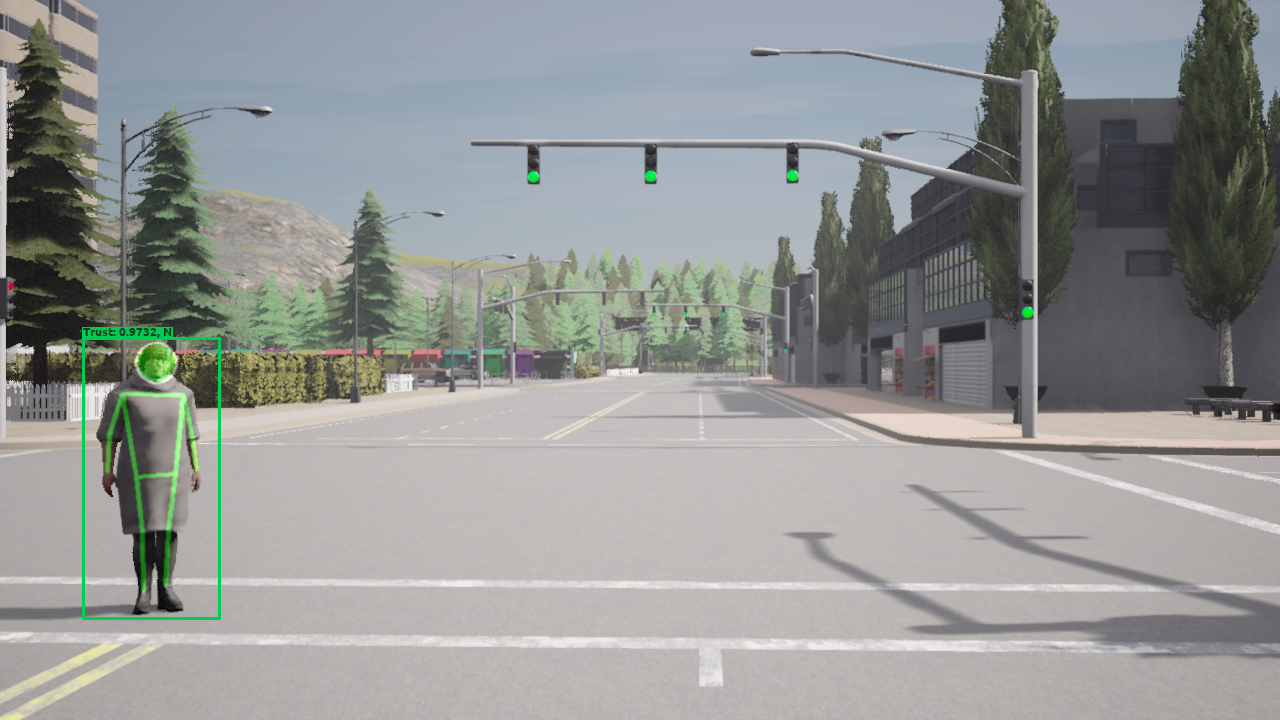}\label{frame4}}
    \caption{Case I: Image frames captured by CARLA camera sensor with visualization scheme for trust estimation. Aggregated trust is monotonically increasing after each image frame that tracks the two pedestrians, with the rate of this increase higher for the attentive pedestrian. The steering input can be observed between Frame 7 and Frame 10, when the steer undertaken by the ego vehicle to introduce extra distance from the distracted pedestrian causes the pedestrian to be out of the field of view of the camera of the ego vehicle. }\label{trust_frames}
  \end{figure*}

To validate the proposed system in an end-to-end manner for real-world applications, simulations were conducted using the CARLA driving simulator \cite{dosovitskiy2017carla} for autonomous driving scenarios. Figure \ref{carla_sim} illustrates the simulated environment, where an ego vehicle navigates a small road segment with simulated pedestrians situated between the ego vehicle and its destination. The ego vehicle is equipped with a camera that captures images of the surrounding environment (refer to Fig.~\ref{trust_frames}). These images are then inputted into the trust estimator module, which computes and aggregates the trust levels of the simulated pedestrians over time. The trust values are incorporated into respective CBF constraints of the MPC controller that governs the behavior of the ego vehicle.

The state model and cost function for the MPC are adapted from \cite{fu_model_predictive_control} which take trajectory information provided by the reference generator to generate control commands (throttle and steer) to the simulator. The system state consists of the position, velocity, orientation of the vehicle along with cross-tracking and orientation errors with respect to the reference path. The optimization problem minimizes these errors along with steer, acceleration, jerk, and the difference in velocity from desired velocity. To ensure robustness for real-world scenarios, a latency of 100 ms is also considered before sending the initial state to the controller. Moreover, it is important to note that although the trust estimator and the optimization problem share the same time step variable $t$ in the formulation presented in (\ref{proposed}), it is impractical to assume they operate at the same frequency. Ideally, the optimization problem would be solved at a much higher sampling rate compared to the trust estimation, due to the computational complexity associated with the image processing algorithms. More details on the system can be seen in the GitHub repository\footnote{https://github.com/saadejazz/mpc-trust-cbf}, along with videos of simulation results.

Two types of pedestrians are simulated as can be seen in Fig.~\ref{ped_type}. One pedestrian maintains a neutral stance and stands at an angle slightly towards the approaching vehicle. Therefore, the trust estimator computes a high trust value for this pedestrian. The other pedestrian's skeleton is manipulated so that the camera suspects smartphone usage - raised hands with bent elbows and neck. Hence, the trust estimator considers this as low trust pedestrian.

The parameters surrounding trust estimation are kept constant throughout the simulations and are as follows: $\rho_1=0.4$, $\rho_2=0.5$, and $\rho_3=0.1$ in (\ref{lin_comb}); $\nu_1=0.6$, and $\nu_{01}=1$ in (\ref{smato_score}); $\nu_2=0.10$ and $\nu_{02}=1$ in (\ref{eye_score}); $\nu_3=0.8$ and $\nu_{03}=0.5$ in (\ref{pose_score}); $\mathcal{F}=0.25$ in (\ref{pose_conf}); $\alpha=1$, $\beta=0.08$, and $\beta_0=0.55$ in (\ref{agg_trust}). Aggregated trust is formulated as monotonically increasing to avoid drastic control commands that can result from conformance to unexpected constraint changes resulting from a sudden decrease in trust. Additionally, the following parameters of the optimization problem are also kept constant throughout the simulations: $R=2.5$, $N_h=7$, $\gamma_{\rm ini}=0.08$, $\delta=0.55$, and $\lambda=2$. Two cases are presented:

\begin{figure}
  \centering
  \includegraphics[width = 0.8\columnwidth]{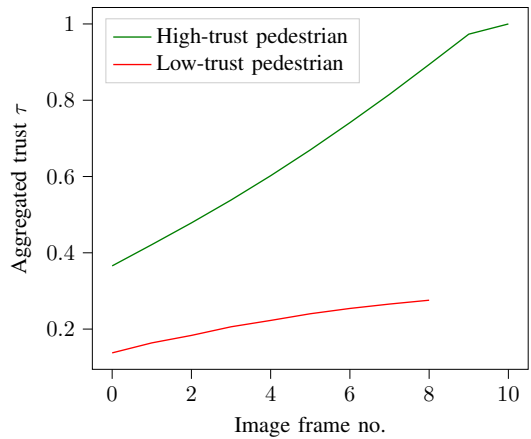}
  \caption{Case I: Dynamics of aggregated trust for pedestrians - the rate of increase of trust of attentive (high-trust) pedestrian is higher than that of the distracted (low-trust) pedestrian}
  \label{trust_scene1}
\end{figure}

\begin{figure}
  \centering
  \includegraphics[width = \columnwidth]{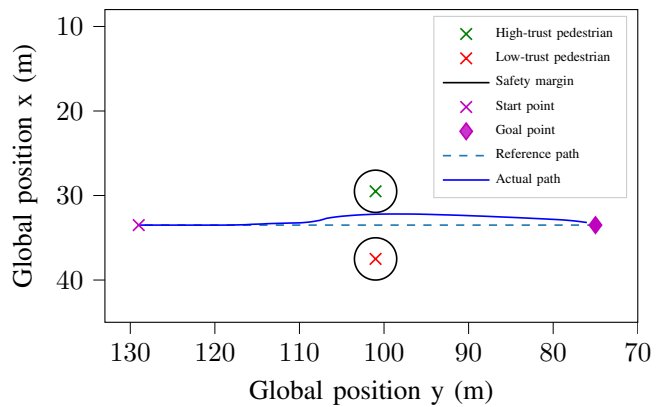}
  \caption{Case I: Path of the ego vehicle - a closer distance is maintained to the the high-trust pedestrian than to the low-trust pedestrian}
  \label{path_carla1}
\end{figure}

\textbf{Case I:} Navigation between an attentive and a distracted stationary pedestrians (see Fig.~\ref{carla_sim})

\textit{Initial Conditions: } $x_e(0) = [33.5, 129]$, $N_p = 2$, $x_1(0) = [29.5, 101]$, $x_2(0) = [37.5, 101]$, $v_1 = [0, 0]$, $v_2 = [0, 0]$, while the goal location of the ego agent is $x_g = [33.5, 75]$.
As the ego vehicle navigates towards its goal, trust is estimated for both pedestrians from captured images. The rate of increase of trust is higher for attentive pedestrians and vice versa. This can be seen in Fig.~\ref{trust_scene1} that depicts the trust levels computed by the trust estimator for each of the pedestrians, from captured images frames shown in Fig.~\ref{trust_frames}. It is important to note that the trust estimation for the pedestrian with low trust is not available for the last two frames. This is because the ego vehicle steered to increase its distance from the inattentive pedestrian and hence that pedestrian was no longer in the field of view of the camera attached to the ego vehicle. This can be observed in Fig.~\ref{frame4}. 

Finally, the path of the ego vehicle can be seen in Fig.~\ref{path_carla1}. The ego vehicle pursued to maintain a larger distance from the untrustworthy pedestrian, as expected. The safety margins of both pedestrians were observed regardless of their trust level.


\textbf{Case II:} Navigation around a moving pedestrian. Two separate simulations are conducted once with the attentive pedestrian and another time with the distracted pedestrian, for the same initial conditions.

\textit{Initial Conditions: } $x_e(0) = [34.5, 125]$, $N_p = 1$, $x_1(0) = [37.5, 101]$, $v_1 = [-0.1, 0]$, while the goal location of the ego agent is $x_g = [34, 75]$.

\begin{figure}
  \centering
  \includegraphics[width = 0.8\columnwidth]{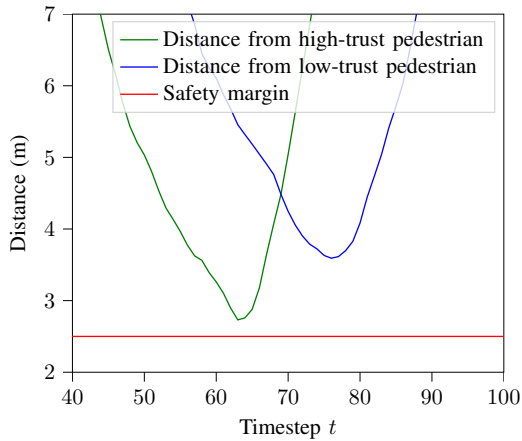}
  \caption{Case II: Distance of ego vehicle from pedestrians - movement around high-trust moving pedestrian is less conservative}
  \label{dist_carla2}
\end{figure}

A distance plot presents an idea of the safety considerations adopted by the ego vehicle. As can be seen in Fig.~\ref{dist_carla2}, the ego vehicle maintained a higher minimum distance from the pedestrian with low trust, when compared to the distance maintained from the pedestrian with high trust.

The results obtained from these simulations provide strong evidence supporting the feasibility of the proposed system, even in the presence of moving pedestrians. However, it is crucial to emphasize that the computational complexity associated with solving the optimization problem, which involves intricate state models and multiple CBF constraints (specifically, $N_p \times N_h$ constraints), in combination with the trust estimation algorithms, can be overwhelming when higher sampling rates are required for real-time applications.
To address this challenge, techniques such as neural network approximation of the optimization problem can be employed to alleviate the computational burden. By approximating the optimization problem using neural networks, the computational complexity can be significantly reduced while maintaining a reasonable level of accuracy. 

\section{Conclusion}

In this paper, we have presented a trust-based safe control system for autonomous navigation in the presence of other agents, specifically human agents or pedestrians. The system combines model predictive control (MPC) with control barrier functions (CBFs) and trust estimation to ensure safe and reliable navigation in complex environments. We have demonstrated the effectiveness of the proposed system through numerical simulations depicting multi-agent systems, and also through autonomous driving simulations in the CARLA driving simulator, where the ego vehicle successfully navigates while considering the trust levels of pedestrians.

The integration of trust estimation into the MPC-CBF framework provides an additional layer of safety by incorporating pedestrian behavior information. The trust estimation algorithm analyzes pedestrian images captured by a camera sensor and computes trust values based on features such as eye contact and smartphone usage. These trust values are then utilized in the CBF constraints of the MPC controller, enabling the ego vehicle to make informed decisions while avoiding potentially risky situations.

The results of our simulations indicate the practicality and feasibility of the proposed system in real-world applications. However, we acknowledge that the computational complexity of the optimization problem and trust estimation algorithms can be challenging for higher sampling rates required in real-time scenarios. To address this issue, we propose future work that focuses on leveraging neural network approximation techniques to reduce the computational burden while maintaining acceptable levels of accuracy. Additionally, further research can be conducted to enhance the trust estimation algorithm by incorporating more behavioural traits to capture a wider range of pedestrian behaviors and improve the accuracy of trust estimation. Moreover, the abundance of parameters used in the proposed framework offer the possibility of applying personalization to adapt to control styles depending on environmental factors of just preference.

In conclusion, the proposed trust-based safe control system offers a promising approach for enhancing the safety and reliability of autonomous navigation in the presence of human. By integrating trust estimation with the MPC-CBF framework, the system can make informed decisions based on human behavior, resulting in safer interactions and mitigating potential risks. Future work focusing on computational efficiency, improved trust estimation, and personalization will further enhance the system's performance and contribute to the development of more sophisticated and reliable autonomous navigation systems.

\bibliographystyle{IEEEtran}
\bibliography{bib}

\begin{thebibliography}{10}
\providecommand{\url}[1]{#1}
\csname url@samestyle\endcsname
\providecommand{\newblock}{\relax}
\providecommand{\bibinfo}[2]{#2}
\providecommand{\BIBentrySTDinterwordspacing}{\spaceskip=0pt\relax}
\providecommand{\BIBentryALTinterwordstretchfactor}{4}
\providecommand{\BIBentryALTinterwordspacing}{\spaceskip=\fontdimen2\font plus
\BIBentryALTinterwordstretchfactor\fontdimen3\font minus
  \fontdimen4\font\relax}
\providecommand{\BIBforeignlanguage}[2]{{%
\expandafter\ifx\csname l@#1\endcsname\relax
\typeout{** WARNING: IEEEtran.bst: No hyphenation pattern has been}%
\typeout{** loaded for the language `#1'. Using the pattern for}%
\typeout{** the default language instead.}%
\else
\language=\csname l@#1\endcsname
\fi
#2}}
\providecommand{\BIBdecl}{\relax}
\BIBdecl

\bibitem{beer2014toward}
J.~M. Beer, A.~D. Fisk, and W.~A. Rogers, ``Toward a framework for levels of
  robot autonomy in human-robot interaction,'' \emph{Journal of Human-Robot
  Interaction}, vol.~3, no.~2, p.~74, 2014.

\bibitem{adams2003trust}
B.~D. Adams, L.~E. Bruyn, S.~Houde, P.~Angelopoulos, K.~Iwasa-Madge, and
  C.~McCann, ``Trust in automated systems,'' \emph{Ministry of National
  Defence}, 2003.

\bibitem{schaefer2013perception}
K.~Schaefer, ``The perception and measurement of human-robot trust,''
  \emph{Ph.D. Thesis}, 2013.

\bibitem{nahavandi2019trust}
S.~Nahavandi, ``Trust in autonomous systems-itrust lab: Future directions for
  analysis of trust with autonomous systems,'' \emph{IEEE Systems, Man, and
  Cybernetics Magazine}, vol.~5, no.~3, pp. 52--59, 2019.

\bibitem{sun2020exploring}
X.~Sun, J.~Li, P.~Tang, S.~Zhou, X.~Peng, H.~N. Li, and Q.~Wang, ``Exploring
  personalised autonomous vehicles to influence user trust,'' \emph{Cognitive
  Computation}, vol.~12, pp. 1170--1186, 2020.

\bibitem{akash2020human}
K.~Akash, G.~McMahon, T.~Reid, and N.~Jain, ``Human trust-based feedback
  control: Dynamically varying automation transparency to optimize
  human-machine interactions,'' \emph{IEEE Control Systems Magazine}, vol.~40,
  no.~6, pp. 98--116, 2020.

\bibitem{akash2019improving}
K.~Akash, K.~Polson, T.~Reid, and N.~Jain, ``Improving human-machine
  collaboration through transparency-based feedback--part i: Human trust and
  workload model,'' \emph{IFAC-PapersOnLine}, vol.~51, no.~34, pp. 315--321,
  2019.

\bibitem{ames2019control}
A.~D. Ames, S.~Coogan, M.~Egerstedt, G.~Notomista, K.~Sreenath, and P.~Tabuada,
  ``Control barrier functions: Theory and applications,'' in \emph{Proceedings
  of the 2019 European Control Conference (ECC)}, 2019, pp. 3420--3431.

\bibitem{ames2017cbfqp}
A.~D. Ames, X.~Xu, J.~W. Grizzle, and P.~Tabuada, ``Control barrier function
  based quadratic programs for safety critical systems,'' \emph{IEEE
  Transactions on Automatic Control}, vol.~62, no.~8, pp. 3861--3876, 2017.

\bibitem{amesadaptive}
A.~D. Ames, J.~W. Grizzle, and P.~Tabuada, ``Control barrier function based
  quadratic programs with application to adaptive cruise control,'' in
  \emph{53rd IEEE Conference on Decision and Control}, 2014, pp. 6271--6278.

\bibitem{sabbir2023trust}
H.~Sabbir~Ahmad, E.~Sabouni, W.~Xiao, C.~G. Cassandras, and W.~Li,
  ``Trust-aware resilient control and coordination of connected and automated
  vehicles,'' \emph{arXiv preprint arXiv:2305.16818}, 2023.

\bibitem{ozkan2022trustaware}
M.~F. Ozkan and Y.~Ma, ``Trust-aware control of automated vehicles in
  car-following interactions with human drivers,'' in \emph{Proceedings of the
  61st IEEE Conference on Decision and Control (CDC)}, 2022, pp. 5279--5284.

\bibitem{zahedi2023trust}
Z.~Zahedi, M.~Verma, S.~Sreedharan, and S.~Kambhampati, ``Trust-aware planning:
  Modeling trust evolution in iterated human-robot interaction,'' in
  \emph{Proceedings of the 2023 ACM/IEEE International Conference on
  Human-Robot Interaction}, 2023, pp. 281--289.

\bibitem{cosner2023learning}
R.~K. Cosner, Y.~Chen, K.~Leung, and M.~Pavone, ``Learning responsibility
  allocations for safe human-robot interaction with applications to autonomous
  driving,'' \emph{arXiv preprint arXiv:2303.03504}, 2023.

\bibitem{lyu2022responsibilityassociated}
Y.~Lyu, W.~Luo, and J.~M. Dolan, ``Responsibility-associated multi-agent
  collision avoidance with social preferences,'' in \emph{Proceedings of the
  2022 IEEE International Conference on Intelligent Transportation Systems
  (ITSC)}, 2022, pp. 3645--3651.

\bibitem{lyu2023risk}
------, ``Risk-aware safe control for decentralized multi-agent systems via
  dynamic responsibility allocation,'' \emph{arXiv preprint arXiv:2305.13467},
  2023.

\bibitem{parwana2022trust}
H.~Parwana, A.~Mustafa, and D.~Panagou, ``Trust-based rate-tunable control
  barrier functions for non-cooperative multi-agent systems,'' in
  \emph{Proceedings of the 61st IEEE Conference on Decision and Control (CDC)},
  2022, pp. 2222--2229.

\bibitem{valtazanos2011intent}
A.~Valtazanos and S.~Ramamoorthy, ``Intent inference and strategic escape in
  multi-robot games with physical limitations and uncertainty,'' in
  \emph{Proceedings of the 2011 IEEE/RSJ International Conference on
  Intelligent Robots and Systems}, 2011, pp. 3679--3685.

\bibitem{brito2021learning}
B.~Brito, A.~Agarwal, and J.~Alonso-Mora, ``Learning interaction-aware guidance
  policies for motion planning in dense traffic scenarios,'' \emph{arXiv
  preprint arXiv:2107.04538}, 2021.

\bibitem{fooladi2021bayesian}
M.~Fooladi~Mahani, L.~Jiang, and Y.~Wang, ``A bayesian trust inference model
  for human-multi-robot teams,'' \emph{International Journal of Social
  Robotics}, vol.~13, no.~8, pp. 1951--1965, 2021.

\bibitem{hu2021trust}
C.~Hu and J.~Wang, ``Trust-based and individualizable adaptive cruise control
  using control barrier function approach with prescribed performance,''
  \emph{IEEE Transactions on Intelligent Transportation Systems}, vol.~23,
  no.~7, pp. 6974--6984, 2021.

\bibitem{frej2022smartphone}
D.~Frej, M.~Ja{\'s}kiewicz, M.~Poliak, and Z.~Zwierzewicz, ``Smartphone use in
  traffic: A pilot study on pedestrian behavior,'' \emph{Applied Sciences},
  vol.~12, no.~24, p. 12676, 2022.

\bibitem{lin2017impact}
M.-I.~B. Lin and Y.-P. Huang, ``The impact of walking while using a smartphone
  on pedestrians’ awareness of roadside events,'' \emph{Accident Analysis \&
  Prevention}, vol. 101, pp. 87--96, 2017.

\bibitem{shinmura2015pedestrian}
F.~Shinmura, Y.~Kawanishi, D.~Deguchi, I.~Ide, H.~Murase, and H.~Fujiyoshi,
  ``Pedestrian's inattention estimation based on recognition of texting while
  walking from in-vehicle camera images,'' \emph{IEICE Technical Report; IEICE
  Tech. Rep.}, vol. 115, no.~98, pp. 83--88, 2015.

\bibitem{shinmura2017recognition}
------, ``Recognition of texting-while-walking by joint features based on arm
  and head poses,'' in \emph{Computer Vision--ACCV 2016, Lecture Notes in
  Computer Science}, 2017, pp. 452--462.

\bibitem{rangesh2016pedestrians}
A.~Rangesh, E.~Ohn-Bar, K.~Yuen, and M.~M. Trivedi, ``Pedestrians and their
  phones-detecting phone-based activities of pedestrians for autonomous
  vehicles,'' in \emph{Proceedings of the 2016 IEEE International Conference on
  Intelligent Transportation Systems (ITSC)}, 2016, pp. 1882--1887.

\bibitem{rangesh2018vehicles}
A.~Rangesh and M.~M. Trivedi, ``When vehicles see pedestrians with phones: A
  multicue framework for recognizing phone-based activities of pedestrians,''
  \emph{IEEE Transactions on Intelligent Vehicles}, vol.~3, no.~2, pp.
  218--227, 2018.

\bibitem{saenz2021detecting}
H.~Saenz, H.~Sun, L.~Wu, X.~Zhou, and H.~Yu, ``Detecting phone-related
  pedestrian distracted behaviours via a two-branch convolutional neural
  network,'' \emph{IET Intelligent Transport Systems}, vol.~15, no.~1, pp.
  147--158, 2021.

\bibitem{hatay2021learning}
E.~Hatay, J.~Ma, H.~Sun, J.~Fang, Z.~Gao, and H.~Yu, ``Learning to detect
  phone-related pedestrian distracted behaviors with synthetic data,'' in
  \emph{Proceedings of the 2021 IEEE/CVF Conference on Computer Vision and
  Pattern Recognition}, 2021, pp. 2981--2989.

\bibitem{uemura2016estimating}
Y.~Uemura, Y.~Kajiwara, and H.~Shimakawa, ``Estimating distracted pedestrian
  from deviated walking considering consumption of working memory,'' in
  \emph{Proceedings of the 2016 International Conference on Computational
  Science and Computational Intelligence (CSCI)}, 2016, pp. 1164--1167.

\bibitem{kusakari2020deep}
Y.~Kusakari, W.~H. Chin, and N.~Kubota, ``A deep neural model for pedestrians
  detection with danger estimation,'' in \emph{Proceedings of the 2020
  International Symposium on Community-centric Systems (CcS)}, 2020, pp. 1--6.

\bibitem{belkada2021pedestrians}
Y.~Belkada, L.~Bertoni, R.~Caristan, T.~Mordan, and A.~Alahi, ``Do pedestrians
  pay attention? eye contact detection in the wild,'' \emph{arXiv preprint
  arXiv:2112.04212}, 2021.

\bibitem{hata2022detection}
R.~Hata, D.~Deguchi, T.~Hirayama, Y.~Kawanishi, and H.~Murase, ``Detection of
  distant eye-contact using spatio-temporal pedestrian skeletons,'' in
  \emph{Proceedings of the 2022 IEEE International Conference on Intelligent
  Transportation Systems (ITSC)}, 2022, pp. 2730--2737.

\bibitem{zhang2018shufflenet}
X.~Zhang, X.~Zhou, M.~Lin, and J.~Sun, ``Shufflenet: An extremely efficient
  convolutional neural network for mobile devices,'' in \emph{Proceedings of
  the 2018 IEEE/CVF Conference on Computer Vision and Pattern Recognition},
  2018, pp. 6848--6856.

\bibitem{kreiss2021openpifpaf}
S.~Kreiss, L.~Bertoni, and A.~Alahi, ``{OpenPifPaf: Composite Fields for
  Semantic Keypoint Detection and Spatio-Temporal Association},'' \emph{IEEE
  Transactions on Intelligent Transportation Systems}, vol.~23, no.~8, pp.
  13\,498--13\,511, 2022.

\bibitem{deng2014pedestrian}
Y.~Deng, P.~Luo, C.~C. Loy, and X.~Tang, ``Pedestrian attribute recognition at
  far distance,'' in \emph{Proceedings of the 22nd ACM international conference
  on Multimedia}, 2014, pp. 789--792.

\bibitem{zheng2017person}
L.~Zheng, H.~Zhang, S.~Sun, M.~Chandraker, Y.~Yang, and Q.~Tian, ``Person
  re-identification in the wild,'' in \emph{Proceedings of the 2017 IEEE/CVF
  Conference on Computer Vision and Pattern Recognition}, 2017, pp. 1367--1376.

\bibitem{wang2007object}
L.~Wang, J.~Shi, G.~Song, and I.-f. Shen, ``Object detection combining
  recognition and segmentation,'' in \emph{Computer Vision--ACCV 2007, Lecture
  Notes in Computer Science}, 2007, pp. 189--199.

\bibitem{cordts2015cityscapes}
M.~Cordts, M.~Omran, S.~Ramos, T.~Scharw{\"a}chter, M.~Enzweiler, R.~Benenson,
  U.~Franke, S.~Roth, and B.~Schiele, ``The cityscapes dataset,'' in
  \emph{Proceedings of the 2015 CVPR Workshop on the Future of Datasets in
  Vision}, vol.~2, 2015.

\bibitem{tan2021efficientnetv2}
M.~Tan and Q.~Le, ``Efficientnetv2: Smaller models and faster training,'' in
  \emph{Proceedings of the 38th International Conference on Machine Learning
  (ICML)}, 2021, pp. 10\,096--10\,106.

\bibitem{zeng2021mpccbf}
J.~Zeng, B.~Zhang, and K.~Sreenath, ``Safety-critical model predictive control
  with discrete-time control barrier function,'' in \emph{Proceedings of the
  2021 American Control Conference (ACC)}, 2021, pp. 3882--3889.

\bibitem{kraft1988software}
D.~Kraft, ``A software package for sequential quadratic programming,''
  \emph{Forschungsbericht- Deutsche Forschungs- und Versuchsanstalt fur Luft-
  und Raumfahrt}, 1988.

\bibitem{dosovitskiy2017carla}
A.~Dosovitskiy, G.~Ros, F.~Codevilla, A.~Lopez, and V.~Koltun, ``{CARLA}: {A}n
  open urban driving simulator,'' in \emph{Proceedings of the 1st Annual
  Conference on Robot Learning}, 2017, pp. 1--16.

\bibitem{fu_model_predictive_control}
J.~Fu, ``Autonomous driving with model predictive control,''
  \url{https://junshengfu.github.io/Model-Predictive-Control/}, accessed: June
  11, 2023.

\end{thebibliography}

\end{document}